\documentclass[useAMS,usenatbib,a4]{mn2e}

\usepackage{times}
\usepackage{graphicx,bm,amssymb}
\usepackage{ulem}
\usepackage{epsfig}
\usepackage{amssymb}
\usepackage{natbib}
\usepackage{psfrag}
\def\araa{ARA\&A}
\def\apj{ApJ}
\def\apjl{ApJ}
\def\apjs{ApJS}
\def\aap{A\&A}
\def\jcap{J. Cosmology Astropart. Phys.}
\def\mnras{MNRAS}
\def\prd{Phys.~Rev.~D}
\def\rmxaa{Rev. Mexicana Astron. Astrofis.}
\def\physrep{Phys.~Rep.}

\newcommand{\be}{\begin{equation}}
\newcommand{\ee}{\end{equation}}
\newcommand{\bary}{\begin{eqnarray}}
\newcommand{\eary}{\end{eqnarray}}

\def\bi{\begin{itemize}}
\def\ei{\end{itemize}}
\def\lsim{\mathrel{\rlap{\lower3pt\hbox{\hskip1pt$\sim$}}
     \raise1pt\hbox{$<$}}} 
\def\gsim{\mathrel{\rlap{\lower3pt\hbox{\hskip1pt$\sim$}}
     \raise1pt\hbox{$>$}}} 

\begin{document}
\title[Signatures of Neutrino Cooling in the SN1987A Scenario]
{Signatures of Neutrino Cooling in the SN1987A Scenario}
\author[N. Fraija et al.]
  {N.~Fraija,$^1$\thanks{E-mail:nifraija@astro.unam.mx. Luc Binette-Fundaci\'on UNAM Fellow.} C.~G. Bernal,$^1$\thanks{E-mail:bernalcg@astro.unam.mx} and  A.~M.~Hidalgo-Gam\'ez$^2$\thanks{E-mail:ahidalgo@esfm.ipn.mx}\\
    $^1$Instituto de Astronom\' ia, Universidad Nacional Aut\'onoma de M\'exico, Circuito Exterior,
C.U., A. Postal 70-264, 04510 M\'exico D.F., M\'exico\\
        $^2$Escuela Superior de F\'isica y Matem\'aticas, Instituto Polit\'ecnico Nacional \\
                U.P. Adolfo L\'opez Mateos, C.P. 07738, M\'exico D.F., M\'exico.}
                
\maketitle
	
\begin{abstract}
The neutrino signal from SN1987A confirmed the core-collapse scenario and the possible formation of a neutron star. Although this compact object has eluded all observations, theoretical and numerical developments have allowed a glimpse of the fate of it. In particular, a hypercritical accretion model has been proposed to forecast the accretion of $\sim 0.15 \:\mathrm{M_{\odot}}$ in two hours and the subsequent submergence of the magnetic field in the newborn neutron star. In this paper, we revisit the Chevalier's model in a numerical framework, focusing on the neutrino cooling effect on the supernova fall-back dynamics.  For that, using a customized version of the FLASH code, we carry out numerical simulations of the accretion of matter onto the newborn neutron star in order to estimate the size of the neutrino-sphere,  the emissivity and luminosity of neutrinos.  As a signature of this phase, we estimate the neutrinos expected on SK neutrino experiment and their flavor ratios.  This is academically important because, although currently it was very difficult to detect 1.46 thermal neutrinos and their oscillations, these fingerprints are the only viable and reliable way to confirm the hypercritical phase. Perhaps new techniques for detecting neutrino oscillations arise in the near future allowing us to confirm our estimates.  
\end{abstract}



\section{Introduction}
It took almost 400 years so that another supernova, after Kepler's supernova in 1604, could be observed by the naked eye. The neutrino signal from SN1987A detected in terrestrial observatories (Mont Blanc, Kamiokande, IBM and Baksan), suggested the formation of a neutron star (NS) by the supernova core \citep{Woosley1988}. So far the search for this holy grail of the type II supernovae paradigm continues.
SN1987A was a core-collapse supernova with known progenitor and an estimated distance of $d\simeq50\pm1\:\mathrm{kpc}$ \citep{Panagia2005}, in the Large Magellanic Cloud (LMC). The progenitor was a blue supergiant of spectral class B3Iab with 20 $\mathrm{M_{\odot}}$, effective temperature $1.6\times10^{4}\:\mathrm{K}$, and an estimated size of 43 $\mathrm{R_{\odot}}$.
The total energy in the collapse was $\simeq10^{53}$ erg, but only $1\%$ of this energy was released in the shock wave of the supernova.\\
In the first few seconds, the debris of core-collapse was dense enough that photons could hardly escape.  However, newly created neutrinos in the formation of the newborn NS  escaped carrying out the gravitational energy. The observables of this NS were  an inferred energy $\sim3\times10^{53}\:\mathrm{ergs}$, temperature $T\sim4\:\mathrm{MeV}$ and decay time-scale of the neutrino burst $\sim4\:\mathrm{s}$. These parameters were consistent with models in which a degenerate iron core collapsed to form a NS \citep{Burrows1986}, although there is not yet any evidence of the presence of a pulsar, or even a quiet NS inside SN1987A. The possible scenarios to solve this problem range from the delayed collapse of the NS into a black hole \citep{Brown-Bethe1994}, a demagnetized NS by hyper-accretion of material \citep{Bernal2013} to the formation of a quark star \citep{Muslimov-Page1995}.\\
In the core-collapse scenario, the shock wave sweeps the outer layers of the progenitor until it encounters a discontinuity in density.  At this point, a reverse shock is generated leading to a fall-back phase which can induce a hypercritical accretion onto the newborn NS surface a few hours after the explosion. This scenario is only possible if the progenitor had a tenuous H/He envelope surrounding a dense He core, as SN1987A \citep{Smartt2009}.  \citet{Chevalier1989} argued in favor of such scenario of late accretion onto the SN1987A core and developed an analytical model for the hypercritical regime. In this model, the neutrino cooling plays an important role in the formation of a quasi-hydrostatic envelope around the newborn NS. Recently,  \cite{Bernal2013, Bernal2010} showed that the magnetic field, for SN1987A parameters, does not play an important role in such regime because it is submerged into the crust of the newborn NS, therefore we will adopt the idea of a demagnetized newborn NS as compact remnant in this supernova. \\
Due to the inverse beta decay, electron-positron annihilation ($e^-+e^+\to Z\to \nu_j+\bar{\nu}_j$) and nucleon-nucleon bremsstrahlung ($N+N\to N+N+\nu_j+\bar{\nu}_j$) for $j=e,\nu,\tau$, thermal neutrinos will be produced at the core and then they will propagate inside the star. The  properties of these neutrinos will get modified when they propagate in a magnetized medium and depending on the flavor of the neutrino, would  feel a different effective potential because electron neutrino ($\nu_e$) interacts with electrons via both neutral and charged currents (CC), whereas muon ($\nu_\mu$) and tau ($\nu_\tau)$ neutrinos interact only via the neutral current (NC). This would induce a coherent effect in which a maximal conversion of $\nu_e$ into $\nu_\mu$ ($\nu_\tau$) takes place even for a small intrinsic mixing angle. The resonant conversion of neutrino from one flavor to another due to the medium effect is well known as the Mikheyev-Smirnov-Wolfenstein effect \citep{wol78}. These  neutrino oscillations have been widely studied in the literature in different scenarios \citep{1999A&A...344..573R,1987ApJ...314L...7G,2000APh....13...21V,2012ARNPS..62..407J,2013MNRAS.434.1355J}.\\
In this work we do a numerical study of hyper-accretion of matter onto the newborn NS surface in the SN1987A scenario. We will consider as input the analytical model proposed by \citet{Chevalier1989} and \citet{Brown-Weingartner1994} about the late accretion of matter onto compact objects, few moments after the core-collapse, focusing on the importance of the neutrino cooling processes on the NS surface. We employ the AMR FLASH code to carry out the numerical simulations of the reverse shock and the complex dynamics near stellar surface, including several neutrino cooling processes, a more detailed equation of state, and an additional degree of freedom in the system. As a signature of this phase we estimate the neutrino luminosity,  the number of events as well as the flavor ratio that could have reached the Earth. The paper is arranged as follows. In section \ref{section2} we describe the analytical model as input on the numerical approach focusing on the neutrino cooling processes present in the system. In section \ref{section3} we develop the neutrino oscillations models. In section \ref{section4} we discuss the radial profiles inside the supernova remnant and the neutrino propagation in such regions. Finally, In section \ref{section5} we present and discuss the results in the SN1987A framework. 
\section{The Hypercritical Accretion Model and the Neutrino Cooling}
\label{section2}
\subsection{The Analytical Procedure}
In the seminal paper about the NS fall-back problem, \citet{Colgate1971} showed that the neutrino cooling on the NS surface would result in a low pressure which could drive matter toward the NS surface. \cite{Chevalier1989} obtained, using this mechanism, that the amount of material deposited on the newborn NS surface, for the SN1987A parameters, was $0.15\:\mathrm{M_{\odot}}$ on a time-scale of $7\times10^{3}$ s. With these parameters, the accretion rate estimated for SN1987A in the hypercritical regime was $\dot{M}\simeq350\:\mathrm{M_{\odot}\, yr.^{-1}}$. He argued that if pressure forces are important in the flow, it has a sonic point at $R_{B}/4$ for $\gamma=4/3$, where $R_{B}$ is the Bondi radius and $\gamma$ is the adiabatic index. Inside of such Bondi flow the velocity becomes supersonic. The conclusion was that the inflow towards the NS surface must be a supersonic free-fall.\\
\citet{Chevalier1989} also found a nice solution that consistently allows the flow to pass through the shock and to decelerate towards the surface of the NS. Additionally, if the infall time is less than the time-scale of the accretion rate change, then the flow can be considered as steady-state. In addition, because of that close to the NS surface the pressure is very high; then it is expected that neutrinos will carry away the gravitational energy. They made several assumptions to find the self-similar solutions: (i) a constant accretion rate and spherical accretion in the hypercritical regime, (ii)  a flow without magnetic field, (iii) they disregard the effects of rotating NS in this regime, (iv) the pair annihilation as the dominant mechanism in the neutrino cooling, and (v) a polytrope approximation to equation of state.\\
Because of that a significant fraction of the material will fall-back onto the compact object (induced by the reverse shock); this material bounces against  the surface of the newly born NS building a third expansive shock, which tries to break through the free falling material. This expansive shock builds an atmosphere (or envelope) in quasi-hydrostatic equilibrium, with free falling  material raining in over it.
In that case, the velocity (v$_0$) and density ($\rho_0$) profiles of the free falling material are given by
\be
v_0=\sqrt{\frac{2GM}{r}}\qquad{\rm and}\qquad \rho_0=\frac{\dot{M}}{4\pi r^{2}v_{0}},
\label{equation1}
\ee
\noindent where $M$ and $\dot{M}$ are the mass and the accretion rate, respectively, and the other parameters have the usual meaning. The structure of the atmosphere in quasi-hydrostatic equilibrium is given by
\be
\rho=\rho_{s}\left(\frac{r_{s}}{r}\right)^{3},\qquad p=p_{s}\left(\frac{r_{s}}{r}\right)^{4},\qquad v=v_{s}\left(\frac{r_{s}}{r}\right)^{-1},
\label{equation2}
\ee
\noindent where the subscript \textit{s} refers to the value of density ($\rho_{s}$), pressure ($p_{s}$) and velocity  ($v_{s}$) at the shock front. The first two values come from imposing hydrostatic equilibrium of a polytropic equation of state, $p\propto\rho^{\gamma}$ and the velocity is fixed by mass conservation.  Once the shock position, $r_{s}$, is known (see below its determination), $p_{s}$ and $\rho_{s}$ are determined by the strong shock condition and $v_{s}$ by mass conservation as
\be
\rho_{s} = 7 \rho_{0},\qquad p_{s}=\frac{6}{7}\rho_{0}v_{0}^{2},\qquad v_{s} = -\frac{1}{7} v_{0},
\label{equation3}
\ee
\noindent where $\rho_{0}$ and $v_{0}$ are evaluated as $r=r_{s}$. For a given NS mass $M$ and radius $R$, and a fixed accretion rate $\dot{M}$, the radial location of the accretion shock is controlled by energy balance between the accretion power and the integrated neutrino losses, per unit NS surface area
\be
\frac{GM\dot{M}}{R} = \int_R^\infty \dot{\epsilon}_{\nu}(r) dr.
\label{equation4}
\ee
\noindent That is because the  value of $r_{s}$ depends on the pressure near the NS surface, where it is attaining a value that is needed to lose the gravitational energy in neutrinos, then the neutrino emissivity, $\dot{\epsilon}_{\nu}(r)$, occurs at the \textit{scale heigh} very close to the NS surface. \citet{Chevalier1989} approached this value by $R/4$ for the pressure profile in an atmosphere in quasi-hydrostatic equilibrium. With this, the energy balance is given by
\be
\frac{GM\dot{M}}{R}=4\pi R^{2}\left(\frac{R}{4}\right)\dot{\epsilon}_{\nu}.
\label{equation5}
\ee 
\noindent The high pressure near the NS surface $(p_{ns}\simeq1.86\times 10^{-12} \mathrm {dyn\, cm^{-2}} \dot{M}\, r_{s}^{3/2})$ allows the pair neutrino process to be the dominant mechanism in the neutrino cooling. The neutrino emissivity can be obtained from \cite{Dicus1972} by
\be
\dot{\epsilon}_{\nu} = 1.83 \times 10^{-34} p^{2.25} \: \mathrm{erg \, cm^{-3} \, s^{-1}} \; .
\label{equation6}
\ee
\noindent Including the electron--positron contribution to the pressure, $p_{e^{-}e^{+}}=11/4\left(aT^{4}/3\right)$ where $a$ is the radiation constant, the shock radius is given by
{\small
\bary\label{rs}
r_{s}&\simeq&7.7\times10^{8}\mathrm{cm} \left(\frac{M}{1.4\,\mathrm{M_{\odot}}}\right)^{-0.04}\left(\frac{R}{10^{6}\mathrm{\, cm}}\right)^{1.48}\cr
&&\hspace{4.0cm}\times\left(\frac{\dot{M}}{\mathrm{M_{\odot}\, yr^{-1}}}\right)^{-0.37}.
\eary
}
\noindent This shock radius is an eingenvalue that allows the gravitational energy to be lost for a given value of $\dot {M}$ which is $r_{s}\simeq8.81\times10^{7}\:\mathrm{cm}$ for SN1987A parameters.
\noindent On the other hand, in the region where the emissivity $\dot{\epsilon}_{\nu}$ is operative, the temperature lies in the range of 1 to 5 MeV, then  the energy density of black body is
\be
B(T)=3.77\times10^{26}\left(\frac{T}{\mathrm{MeV}}\right)^4\:\mathrm{erg\, cm^{-3}},
\label{equation8}
\ee
\noindent and the neutrino emissivity as a function of temperature, in the hypercritical regime is calculated as
\be
\dot{\epsilon}_{\nu}=0.97\times10^{25}\left(\frac{T}{\mathrm{MeV}}\right)^{9}\:\mathrm{erg\, s^{-1}\, cm^{-3}},
\label{equation9}
\ee 
where we have taken into account  the contribution of  e$^\pm$ pairs.\\
\noindent For the SN1987A parameters, the temperature estimated  on the NS surface is $T\simeq4.5\times10^{10} \, \mathrm{K} \: (\simeq4 \, \mathrm{MeV})$,  and the emissivity is given by $\dot{\epsilon}_{\nu}\simeq2\times10^{30}\:\mathrm{erg\, s^{-1}\, cm^{-3}}$. 
As the volume of the neutrino-sphere is $V\simeq\pi R^{3}=\pi\times10^{18}\:\mathrm{cm^{3}}$, the neutrino luminosity is given by $L_{\nu}\simeq6\times10^{48}\:\mathrm{erg\, s^{-1}}$. Note that although in this case other processes of neutrino production are neglected due to the high dependence of the pair annihilation process with temperature, in the numerical approach, we will include all the relevant processes.
\subsection{The Numerical Technique}
\begin{figure*}[htp]
\centering
\includegraphics[width=\textwidth]{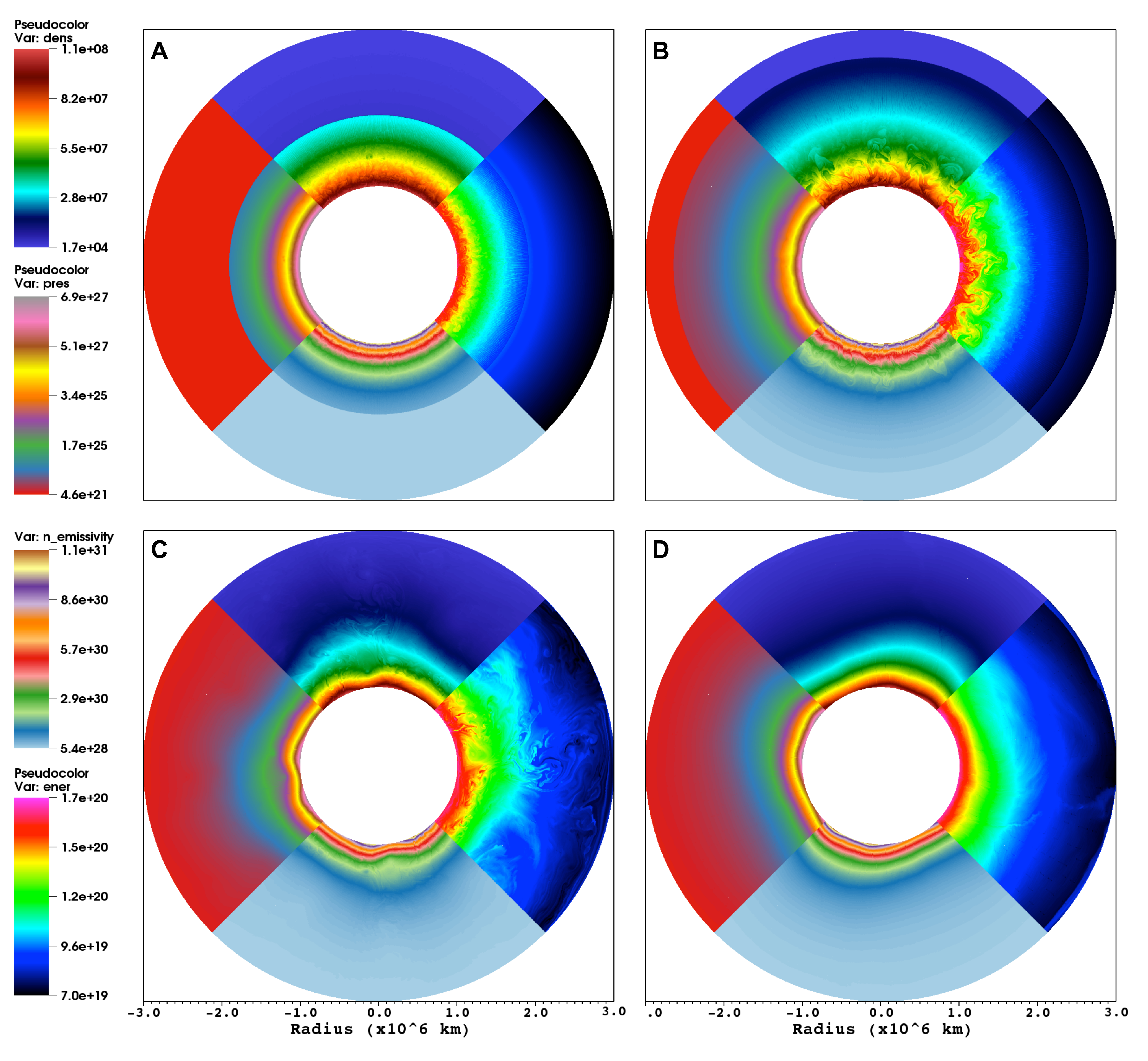}
\caption{\label{Figure1} Color maps of the evolution of density (up), total energy (right), neutrino emissivity (down) and pressure (left) for the SN1987A parameters: (A) initial transient, $t=0.1$ ms, (B) shock evolving in the domain, $t=0.5$ ms, (C) shock leaves the domain and the transient has vanished, $t=5$ ms, (D) quasi-hydrostatic  equilibrium envelope is formed, $t=10$ ms.}
\end{figure*}

\begin{figure*}[htp]
\centering
\includegraphics[width=\textwidth]{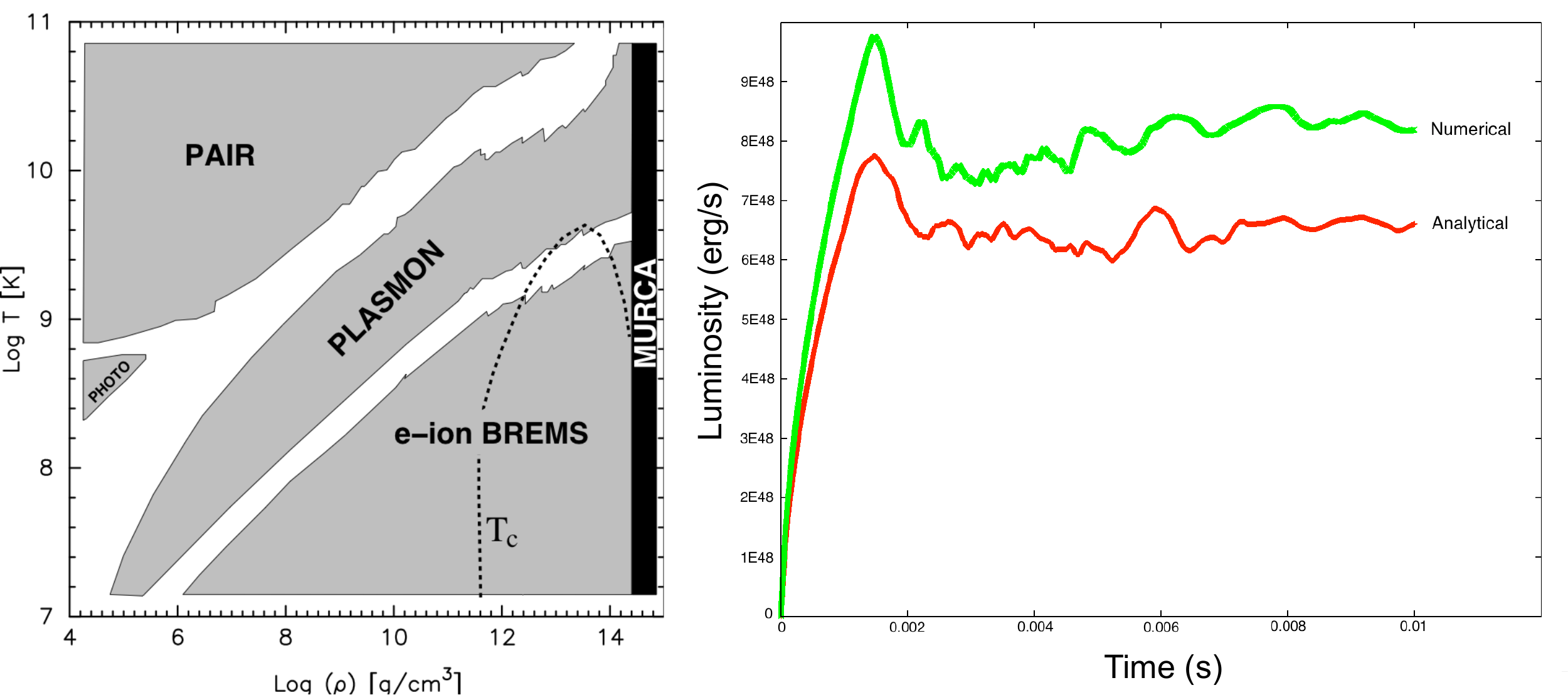}
\caption{\label{Figure2} Left: Domain of validity of the neutrino cooling processes in the density-temperature space. We used the Itoh tabulated values for several neutrino processes \citep{Itoh1996}. Each shaded region shows the regime in which a given process contributes more than 90\% of the neutrino energy losses. As a comparative example the neutron $^1S_0$ superfluidity $T_c$ curve is also plotted (dotted line), below which the PBF (Cooper pair breaking and formation) process starts to act with an emissivity similar to the one of the e-ion bremsstrahlung. Right: neutrino luminosity integrated in the whole computational domain. The analytical formula from Dicus \citep{Dicus1972} and the tabulated numerical values from Itoh \citep{Itoh1996} are compared.}
\end{figure*}

Although the simplified Chevalier model described above estimates the values of parameters and the radial structure of the atmosphere in quasi-hydrostatic equilibrium at the hypercritical regime, it is only a one-dimensional model and also requires several assumptions as was pointed out: polytrope approximation for the equation of state, just one process for the neutrino cooling (pair production) and negligible magnetic field, for instance.\\
\citet{Bernal2013, Bernal2010} showed that the magnetic field, for SN1987A parameters, is buried and submerged under the stellar surface by the accreting material in the hypercritical regime, so it does not play an important roll in the dynamics of quasi-hydrostatic envelope. Because of that the ram pressure is greater than the magnetic pressure in this regime, the magnetic field is confined in a small region where the piling up of matter takes place. In such cases, numerical simulations of cartesian two-three-dimensional accretion columns must be considered in order to take into account the role of the magnetic field. The results show the piling up of matter onto the NS surface, as well as the submergence of the magnetic field in the new crust formed by such material.
For such reasons we do not take into account the magnetic field in the present work. Otherwise, we will include more neutrino processes, a more detailed equation of state and one additional degree of freedom in the simulations in order to compute  several parameters of the neutrino cooling effect on the stellar surface, an instance after the material has fallen back.\\
\noindent In the present case, we carry out hydrodynamic numerical simulations of the hypercritical accretion regime in a two-dimensional spherical mesh, using the Flash code method developed by  \citet{Fryxell2000}. Flash is a Eulerian, parallelized, multi-physics, adaptive mesh code designed to handle several problems found in various high-energy astrophysical environments. In our case, we use a customized version of Flash code, with the piecewise--parabolic method PPM solver, which solves a whole set of hydrodynamic equations. This solver uses an algorithm which is a version of higher order Godunov's scheme.  The matter equation of state is an adaptation of Flash's Helmholtz package, that includes contributions from the nuclei, electron-positron pairs, and radiation, as well as the Coulomb correction. \\
\noindent The neutrino energy losses near the stellar surface are dominated by the $e^\pm$ annihilation process which involves the formation of neutrino-antineutrino pairs  ($e^{-}+e^{+}\rightarrow\nu+\bar{\nu}$).  However, we also include other relevant neutrino processes present in such regime: (i) the photo-neutrino process, in which the outgoing photon in a Compton scattering is replaced by a neutrino-antineutrino pair ($\gamma+e^{\pm}\rightarrow e^{\pm}+\nu+\bar{\nu}$); (ii) the plasmon decay process, in which a photon propagating within an electron gas (plasmon) is spontaneously transformed in to a neutrino-antineutrino pair ($\gamma\rightarrow\nu+\bar{\nu}$) and (iii) the Bremsstrahlung process, in which the photon of the standard process is replaced by a neutrino-antineutrino pair, either due to electron-nucleon interactions ($e^{\pm}+N\rightarrow e^{\pm}+N+\nu+\bar{\nu}$) or nucleon-nucleon interactions ($N+N\rightarrow N+N+\nu+\bar{\nu}$). All these processes, which are implemented in a customized module in the code,  are described in \cite{Itoh1996}. \\
\noindent We used a 2D spherical mesh $\left(r,\theta\right)$ to perform the numerical simulations. The radial component $r$ lies in the range $10^{6} {\rm cm} \leq r\leq 3\times10^{6} {\rm cm}$, and  the angle $\theta$  in the range $\pi/4\leq\theta\leq3\pi/4$, i.e, we simulated only a quarter of the total domain, which has $2048\times2048$ effective zones.\\
\noindent As boundary conditions, we impose mass inflow along the top edge of the computational domain, and periodic conditions along the sides. At the bottom, on the NS surface, we use a custom boundary condition that enforces hydrostatic equilibrium. We assume the accreting matter to be non-magnetized. We are interested in following the evolution of the system from the  instant when the material in free fall bounces against the NS surface. \citet{Bernal2010, Bernal2013} showed that the radial profiles for density, pressure and velocity described in equation \ref{equation2} are achieved when the system evolves a long time. They found, besides the aforementioned profiles predicted by the analytical model, a submergence of the NS magnetic field in the new crust formed by the material piled on the stellar surface.\\
In this work we are not interested in following the shock evolution and the consequent formation of a quasi-hydrostatic equilibrium envelope, but we focus on the dynamics very close to the NS surface where the neutrino cooling takes place. The boundary condition on the top edge is adaptable, i.e, when the shock leaves the computational domain, the injection of mass changes from free fall to Chevalier mode. We start the simulation with the free fall profiles described by eq. \ref{equation1} with an initial temperature of $10^{7}$ K. The code finds the correct pressure profile after some steps of simulation.
The time step in the Flash code is adaptive and depends on local conditions. Typically, the time resolution of the simulations is $dt\simeq10^{-7}$~s. 
It is important to highlight that once the atmosphere in quasi-hydrostatic equilibrium is formed, it will oscillate around the shock radius. These oscillations will depend on the physical conditions on NS surface as well as the amount of neutrinos that  carry away the energy injected by the accretion.  With the present conditions of density and temperature at the base of the flow, the plasma is made of  free baryons, photons and e$^\pm$. The main contributions of neutrino opacity are then the coherent scattering of neutrons and protons and pair annihilation.   For example, the corresponding cross section for coherent scattering is $\sigma_{\rm N}=(1/4)\sigma_0[E_{\nu}/(m_e c^2)]^2$, where $\sigma_0=1.76 \times 10^{-44}$~cm$^{2}$. 
For thermal neutrinos  with temperatures  $T \lesssim 10^{11}$ K $\sim 10$ MeV, the cross section is $\sigma_{\rm N} \lesssim 7 \times 10^{-42}$ cm$^2$. The maximum densities reached at the bottom of the envelope will be less than $10^{11}$ g cm$^{-3}$, and in such conditions, the neutrino mean free-path would be $l_{\nu}=(n_{\rm N} \sigma_{\rm N})^{-1} \gtrsim 2.5 \times 10^{6}$~cm, which is safely larger than the depth of the dense envelope. Above this dense region the envelope density decreases rapidly and the whole envelope is transparent to neutrinos. Then, we ignore neutrino absorption and heating. Also, if we ignore convection effects, the time scale required for the quasi-stationary solution to set in is a few sound crossing times, $t_{\rm cross}\simeq r_{\rm s}/c_{\rm s}$.  For a shock radius of $r_{\rm s}\simeq 50$~km and sound velocity $c_{\rm s}\simeq c/10$, this is $t_{\rm cross}\simeq 1-2$~ms. The simulations presented by \citet{Bernal2013, Bernal2010} ran for hundreds of ms in order to  establish itself quite rapidly. In the present case,  to analyze the initial transient and the neutrino cooling effect over the stellar surface, our simulations ran for various ms. Because of that the neutrino cooling depends on temperature, being more or less constant near the NS surface; the energy loss has the same behavior in the hypercritical regime where such emissivity is operative. The time scale for convection is of course much longer, and will depend on the equilibrium between infall and cooling at the base of the envelope.\\
Fig. \ref{Figure1} shows color maps of density (up), total energy (right), emissivity (down) and pressure (left) for the SN1987A parameters. In (A) we show the initial transient,  with free-falling material bouncing on the stellar surface and building an expansive shock, that makes its way through the material falling onto the remnant. It is important to highlight that the pressure and energy are high pretty close to the NS surface where the neutrino cooling takes place ($t=0.1$ ms). In (B) the shock evolution is evident and a rich morphology is observed in the system. Hydrodynamic instabilities are observed inside the envelope while the shock evolves in the computational domain ($t=0.5$ ms). In (C) the shock has left the computational domain while at the base of the envelope the neutrino cooling is very effective creating an energy sink that allows the material to be deposited on the surface slowly. At this stage the system begins to relax ($t=5$ ms). In (D) the system has nearly reached a quasi-hydrostatic equilibrium. Because of  the additional degree of freedom, the flow passes freely through the lateral boundaries, preventing a full equilibrium state from reaching. Nevertheless, it can be observed that the height scale where neutrino-sphere is operating,  the emissivity is more efficient.\\
\noindent From the simulations it has been computed that  the estimated radius where the neutrino loss is effective (including all the relevant processes) is $r\simeq3.2\times10^{5}\:\mathrm{cm}\simeq(1/3)R$. The mean value of emissivity in such region is $\dot{\epsilon}_{\nu}\simeq2.2\times10^{30}\:\mathrm{erg\, s^{-1}\, cm^{-3}}$ (very similar to the one  analytically estimated ), the volume of the neutrino-sphere is $V\simeq(4/3)\pi R^{3}=(4/3)\pi\times10^{18}\:\mathrm{cm^{3}}$ and then the neutrino luminosity is given by $L_{\nu}\simeq8\times10^{48}\:\mathrm{erg\, s^{-1}}$.  The good agreement of these results with the estimated values of the analytical model is surprising.\\
In Fig. \ref{Figure2} is shown, the parameter space where different neutrino processes are effective as well as the neutrino luminosity integrated in the whole computational domain, for the SN1987A parameters. Note that after an initial transient, the neutrino luminosity has small oscillations about a fixed value. Two approaches are shown, the analytical method from \citet{Dicus1972} and the tabulated method from \citet{Itoh1996}. The luminosity values are in perfect agreement with those calculated analytically above. We can infer that the small difference between the analytical and numerical values calculated by integrating the whole computational domain is due to other neutrino production processes not taken into account in the analytical model. We confirm that this nice analytical model accounts for many important physical processes in the hypercritical regime. However, numerical simulations allow us to go beyond and analyze other relevant physical processes that are lost in purely analytical models, such as the magnetic field submergence, the piling up of matter onto the NS surface and the rich hydrodynamic morphology and instabilities that only numerical simulations can reproduce.\\
Notably, the numerical values obtained by simulations are estimated once the system has reached a quasi-hydrodynamic equilibrium. In this case, most of instabilities presented in the initial transient has disappeared, thus allowing to make a comparative analysis with the analytical model.\\
\subsection{Number of   Expected Neutrinos}
For the hypercritical phase we have calculated the number of events that could have been detected in SKII. The expected event rate can be written as 
\be\label{rate}
N_{ev}=V N_A\,  \rho_N \int_t \int_{E'} \sigma^{\bar{\nu}_ep}_{cc} \frac{dN}{dE}\,dE\, dT
\ee

\noindent where $V$ is the effective volume of water, $N_A=6.022\times 10^{23}$ g$^{-1}$ is the Avogadro's number, $\rho_N=2/18\, {\rm g\, cm^{-3}}$ is the nucleons density in water \citep{2004mnpa.book.....M}, $ \sigma^{\bar{\nu}_ep}_{cc}\simeq 9\times 10^{-44}\,E^2_{\bar{\nu}_e}/MeV^2$  is the cross section \citep{1989neas.book.....B}, $dT$ is the detection time of neutrinos  and $dN/dE$ is the neutrino spectrum. Taking into account the relationship between the  neutrino luminosity $L$ and neutrino flux $F$, $L=4\pi D^2_z  F<E>=4\pi D^2_z   E^2 dN/dE$ and approximation of the time-integrate average energy  $<E_{\bar{\nu}_e}>=15$ MeV and time $t$, then the number of events (eq. \ref{rate}) are
\bary
N_{ev}&\simeq&\frac{t}{<E_{\bar{\nu}_e}>}V N_A\,  \rho_N  \sigma^{\bar{\nu}_ep}_{cc} <E_{\bar{\nu}_e}>^2\frac{dN}{dE}\cr
&\simeq&\frac{t}{4\pi D^2_z <E_{\bar{\nu}_e}>}V N_A\,  \rho_N  \sigma^{\bar{\nu}_ep}_{cc}\,L_{\bar{\nu}_e.}
\eary
\noindent Replacing the values for a water volume of $V=2.14\times 10^{9}\, {\rm cm}^{3}$ (2.14 kton) \citep{2004mnpa.book.....M}  and neutrino luminosity obtained $L_\nu \simeq \left(8.0/6.0\right) \times 10^{48}$  erg/s,   we have that the number of neutrinos expected  would have been 1.49  which is at the limit of detection and in comparison to the value of initial neutrino burst 18.7 \citep{2007fnpa.book.....G,2004mnpa.book.....M}  is very small.\\
On the other hand, when such neutrinos are produced in the neutrino-sphere, they can have a very complex behavior.
In the following sections, we calculate the neutrino effective potential and the resonant oscillations of these neutrinos since they are produced in the neutrino-sphere until they go through the upper layers of the supernova progenitor and reach the Earth.

\section{Neutrino Oscillations}
\label{section3}

The  properties of these  neutrinos are modified when they propagate in such magnetized and thermal medium and  depending on the neutrino flavor, it would  feel a different effective potential because electron neutrino ($\nu_e$) interacts with electrons via both neutral and charged currents (CC), whereas muon ($\nu_\mu$) and tau ($\nu_\tau)$ neutrinos interact only via the neutral current (NC). This would induce a coherent effect in which maximal conversion of $\nu_e$ into $\nu_\mu$ ($\nu_\tau$) takes place even for a small intrinsic mixing angle \citep{wol78}.   These neutrino oscillations have been widely studied in the literature in different scenarios \citep{2000APh....13...21V, 1987ApJ...314L...7G, 1999A&A...344..573R, 2005PhRvD..71d7303S, 2009PhRvD..80c3009S, 2009JCAP...11..024S, 2013arXiv1304.4906O,2014arXiv1401.1908F, 2014arXiv1401.3787F,2014arXiv1401.4615F}.   In the following subsections we are going to introduce the more relevant equations for neutrino oscillations in matter that are derived  in \cite{2014arXiv1401.1581F}. 
\subsection{Two-Neutrino Mixing}
The evolution equation for the propagation of neutrinos in matter is given by
\be
i
{\pmatrix{
\dot{\nu}_{e} \cr
\dot{\nu}_{\mu}
}}
=
{\pmatrix{
V_{eff}-\Delta \cos 2\theta & \frac{\Delta}{2}\sin 2\theta \cr
\frac{\Delta}{2}\sin 2\theta  & 0
}}
{\pmatrix{
\nu_{e} \cr
\nu_{\mu}
}}
\ee
where $\Delta=\delta m^2/2E_{\nu}$, $V_{eff}$ is the effective potential,   $E_{\nu}$ is the neutrino energy,  $\delta m^2$ is the neutrino mass difference and $\theta$ is the neutrino mixing angle.  Here we have considered the neutrino oscillation process $\nu_e\leftrightarrow \nu_{\mu, \tau}$.   From  the conversion probability $P_{\nu_e\rightarrow {\nu_{\mu}{(\nu_\tau)}}}(t) = \Delta^2 \sin^2 2\theta/\omega^2\,\sin^2\left (\omega t/2\right)$ with $\omega=\sqrt{(V_{eff}-\Delta \cos 2\theta)^2+\Delta^2 \sin^2 2\theta}$, the oscillation length for the neutrino is written as
\be
L_{osc}=\frac{L_v}{\sqrt{\cos^2 2\theta (1-\frac{V_{eff}}{\Delta \cos 2\theta})^2+\sin^2 2\theta}},
\label{osclength}
\ee
where $L_v=2\pi/\Delta$ is the vacuum oscillation length. If the density of the medium is such that the condition $V_{eff}=\Delta\,\cos2\theta$ is satisfied, then the resonance condition and resonance length can be written as
\be
V_{eff}=\Delta \cos 2\theta\,,
\label{reso2d}
\ee
and%
\be
L_{res}=\frac{L_v}{\sin 2\theta},
\label{oscres}
\ee
respectively. Combining eqs. (\ref{reso2d}) and (\ref{oscres}),  we can obtain the resonance density as a function of resonance length  \citep{2014MNRAS.437.2187F}
{\scriptsize 
\begin{equation}
\textbf{$\rho_R$}=
\cases{
\frac{3.69\times 10^{2}}{E_{\nu,MeV}}    \,     \biggl[ 1- E_{\nu,MeV}^2\biggl( \frac{4.4 \times 10^{6} \,cm}{l_r}\biggr)^2  \biggr]^{1/2}{\rm gr/cm^3}       &   {\rm sol}\,,  \nonumber\cr
\frac{1.39\times 10^{4}}{E_{\nu,MeV}}    \,     \biggl[ 1- E_{\nu,MeV}^2\biggl( \frac{1.18 \times 10^{5}\,cm}{l_r}\biggr)^2  \biggr]^{1/2} {\rm gr/cm^3}     &   {\rm atm.}\,,\nonumber\cr
\frac{3.29\times 10^{6}}{E_{\nu,MeV}}    \,     \biggl[ 1- E_{\nu,MeV}^2\biggl( \frac{4.9 \times 10^2\,cm}{l_r}\biggr)^2  \biggr]^{1/2} {\rm gr/cm^3}          &   {\rm acc.}\,. \cr
}
\label{p1}
\end{equation}
}
We will be using the best fit values of two-neutrino mixing (solar, atmospheric and accelerator neutrino experiments) as follows.   Here we will consider the following neutrino mixing parameters. From {\bf solar neutrinos}:  $\delta m^2=(5.6^{+1.9}_{-1.4})\times 10^{-5}\,{\rm eV^2}$ and $\tan^2\theta=0.427^{+0.033}_{-0.029}$\citep{aha11},  from  {\bf atmospheric neutrinos}:  $\delta m^2=(2.1^{+0.9}_{-0.4})\times 10^{-3}\,{\rm eV^2}$ and $\sin^22\theta=1.0^{+0.00}_{-0.07}$ \citep{abe11a} and from  {\bf accelerator neutrinos}:  $\delta m^2=(7.9^{+0.6}_{-0.5})\times 10^{-5}\,{\rm eV^2}$ and $\tan^2\theta=0.4^{+0.10}_{-0.07}$\citep{ara05,shi07,mit11}. 
\subsection{Three-neutrino Mixing}
In a three-flavor framework, the evolution equation is given by
\be
i\frac{d\vec{\nu}}{dt}=H\vec{\nu},
\ee
where the state vector in the flavor basis is $\vec{\nu}\equiv(\nu_e,\nu_\mu,\nu_\tau)^T$, the effective Hamiltonian is $H=U\cdot H^d_0\cdot U^\dagger+diag(V_{eff},0,0)$, with $H^d_0=\frac{1}{2E_\nu}diag(-\Delta m^2_{21},0,\Delta^2_{32})$ and $U$  the three neutrino  mixing matrix given by \cite{gon03,akh04,gon08,gon11}. The survival and conversion  probabilities  for electron, muon and tau are given in \cite{2009JCAP...11..024S, 2009PhRvD..80c3009S, 2014arXiv1401.1581F}. The oscillation length for the neutrino is given by
\be
L_{osc}=\frac{L_v}{\sqrt{\cos^2 2\theta_{13} (1-\frac{2 E_{\nu} V_e}{\delta m^2_{32} \cos 2\theta_{13}} )^2+\sin^2 2\theta_{13}}},
\label{osclength}
\ee
where $L_v=4\pi E_{\nu}/\delta m^2_{32}$ is the vacuum oscillation length. From the resonance condition,  $V_{eff}=\Delta \cos2\theta_{13}$, the resonance length and density  are related as 
\be 
\rho_R=\frac{1.9\times 10^{4}}{E_{\nu,MeV}}    \,     \biggl[ 1- E_{\nu,MeV}^2\biggl( \frac{8.2 \times 10^{4} \,cm}{l_r}\biggr)^2  \biggr]^{1/2}{\rm gr/cm^3}\,.
\label{p2}
\ee
On the other hand, combining solar, atmospheric, reactor and accelerator parameters, the best fit values of the three neutrino mixing  are ${\rm for}\,\,\sin^2_{13} < 0.053:  \Delta m_{21}^2= (7.41^{+0.21}_{-0.19})\times 10^{-5}\,{\rm eV^2};   \hspace{0.1cm} \tan^2\theta_{12}=0.446^{+0.030}_{-0.029}$ and ${\rm for}\,\,\sin^2_{13} < 0.053:\Delta m_{21}^2= (7.41^{+0.21}_{-0.19})\times 10^{-5}\,{\rm eV^2};   \hspace{0.1cm} \tan^2\theta_{12}=0.446^{+0.030}_{-0.029}$\citep{aha11}.
\subsection{Neutrino Oscillation from  Source to Earth}
Between the surface of the star and the Earth the flavor ratio $\phi^0_{\nu_e}:\phi^0_{\nu_\mu}:\phi^0_{\nu_\tau}$   is affected by the full three description  flavor mixing. The probability for a neutrino to oscillate from a flavor estate $\alpha$ to a flavor state $\beta$ in a time starting from the emission of neutrino at star $t=0$, is given as {\small $P_{\nu_\alpha\to\nu_\beta} =\mid <  \nu_\beta(t) | \nu_\alpha(t=0) >  \mid =\delta_{\alpha\beta}-4 \sum_{j>i}\,U_{\alpha i}U_{\beta i}U_{\alpha j}U_{\beta i}\,\sin^2\left(\delta m^2_{ij} L/4\, E_\nu \right)$}.  With the three-mixing parameters of neutrino oscillation,  we can write the mixing matrix as
\be
U =
{\pmatrix{
0.82      &  0.55     &     0.19\cr
 -0.51  & 0.51      &     0.69\cr
 0.28   &  -0.66    &     0.69
}}
\ee
and the probability matrix for a neutrino flavor vector of ($\nu_e$, $\nu_\mu$, $\nu_\tau$)$_{source}$ changing to a flavor vector  ($\nu_e$, $\nu_\mu$, $\nu_\tau$)$_{Earth}$ is given as
\be
{\pmatrix{
\nu_e   \cr
\nu_\mu   \cr
\nu_\tau   
}_{Earth}}
=
{\pmatrix{
0.53     & 0.27      & 0.20\cr
0.27      & 0.37      &  0.37\cr
0.20      & 0.37      & 0.43
}}
{\pmatrix{
\nu_e  \cr
\nu_\mu \cr
\nu_\tau 
}_{source}}
\label{matrixosc}
\ee
where  the sin term in the probability has been averaged  to $\sim 0.5$ for distances (L) longer than the Solar System. \citep{lea95}. 
\section{Density Profiles} 
\label{section4}
\begin{figure*}[htp]
\centering
\includegraphics[width=\textwidth]{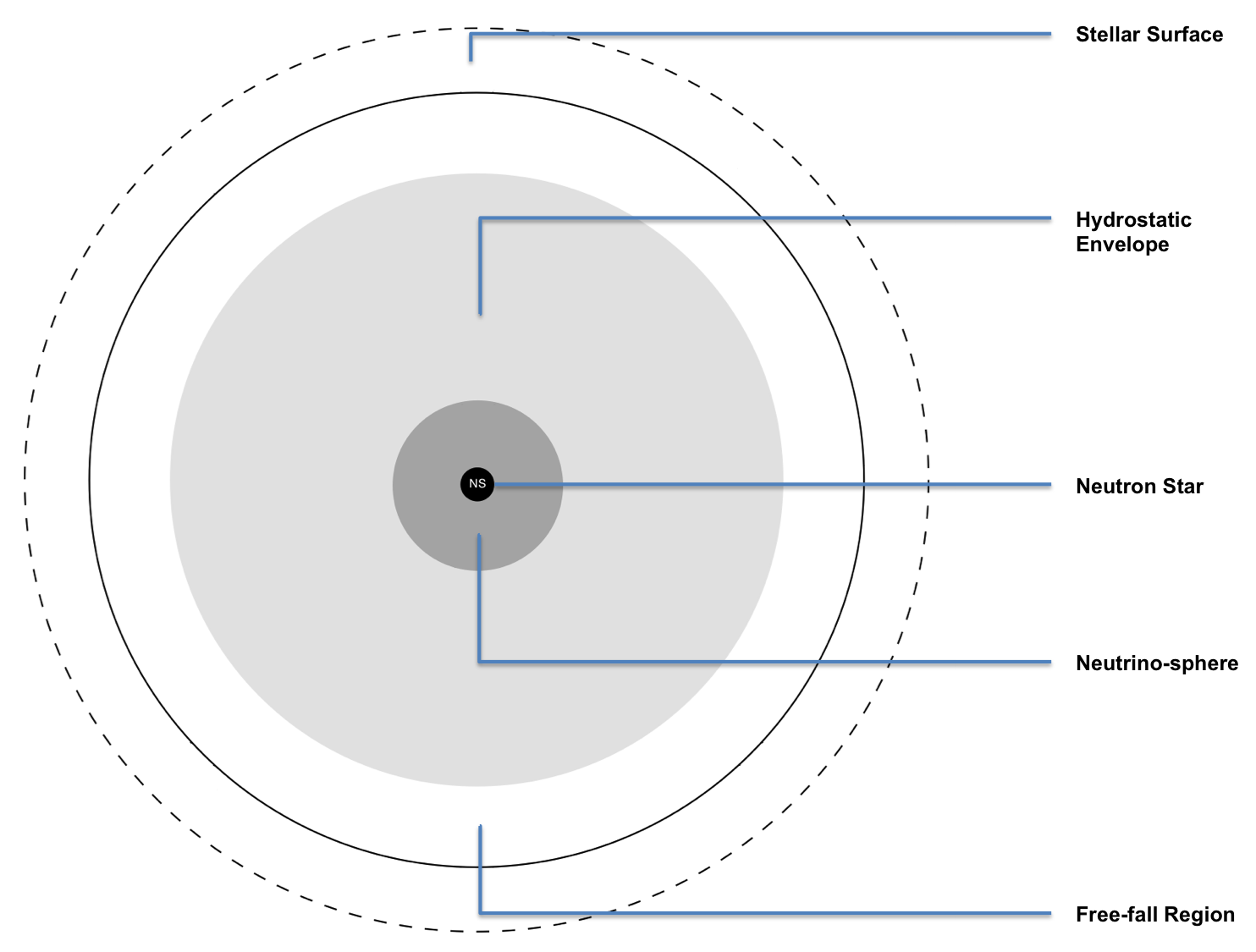}
\caption{\label{Figure3} Important regions inside early supernova remnant: the newborn NS, the core surface (neutrino-sphere), the hydrostatic envelope, the free-fall region and the stellar surface. The neutrinos produced in the neutrino-sphere travel through these region oscillating.}
\end{figure*}

When neutrinos propagate inside a star they have to go through different stratified regions. In order to obtain the density profiles, we have selected four important regions inside the star (see Fig. \ref{Figure3}). The core surface (neutrino-sphere), the accretion shock region (hydrostatic envelope), the free-fall region and the stellar surface. Analytical and numerical models of density distribution in a pre-supernova have shown a decreasing dependence on radius $\rho\propto r^{-n}$, with $n=3/2 \, (3)$ for convective (radiative) envelopes \citep{woo93, shi90, arn91}.
 \subsection{Core Surface}
\begin{figure*}[htp]
\centering
 \includegraphics[width=\textwidth]{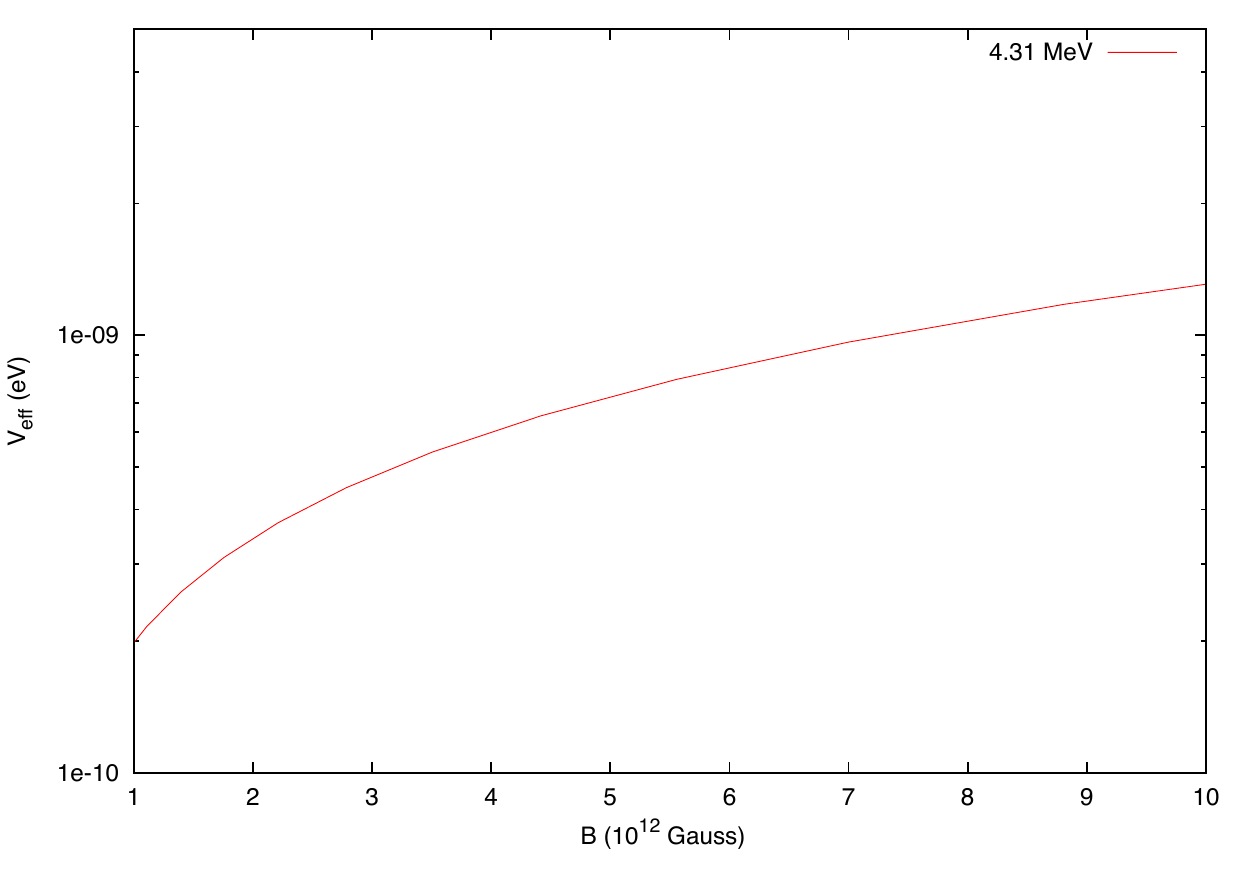} 
\caption{\label{Veff}  The effective potential (V$_{eff}$) as a function of magnetic field (B) is plotted for  4.31 MeV.}
\end{figure*}

\begin{figure*}[htp]
\centering
\includegraphics[width=\textwidth]{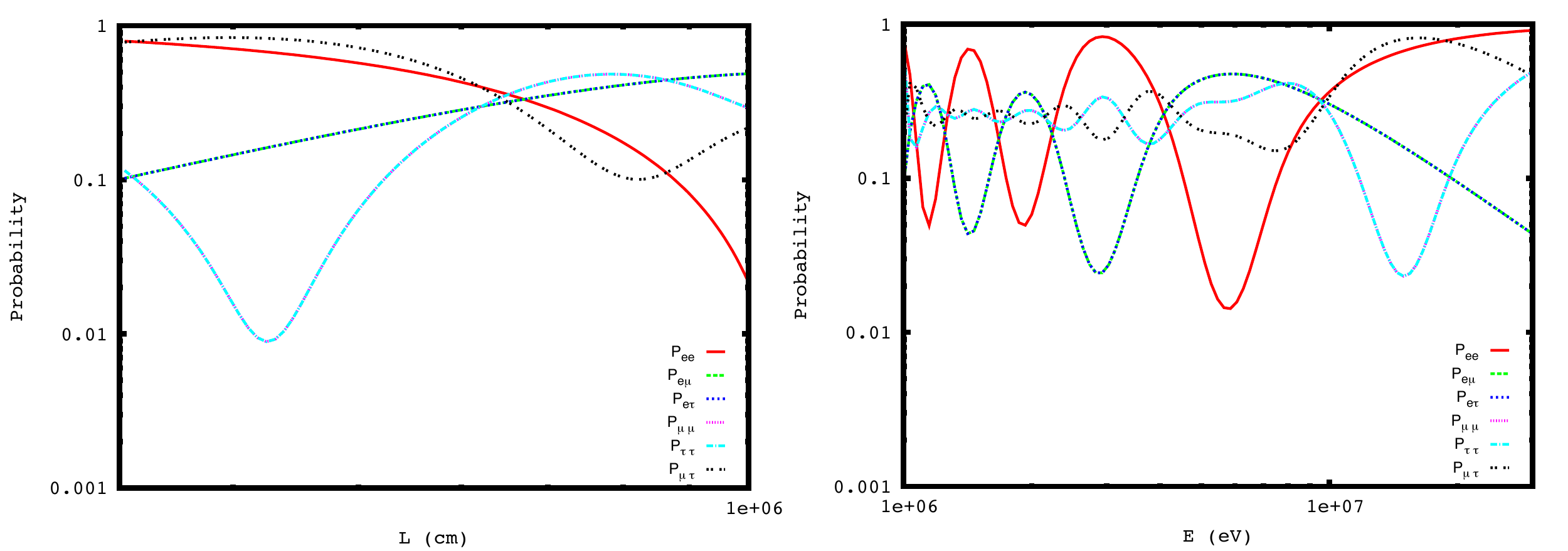}
\caption{\label{pr_r1} We plot the oscillation probability as a function of distance (left figure) and energy (right figure) when neutrinos are propagating through region 1.}
\end{figure*}

On the surface of the core,  the plasma is magnetized and thermal  (i.e. the neutrino-sphere). In this region ($1.0\times10^6 {\rm cm}  \leq r_1\leq 1.33\times10^6 {\rm cm}$), the magnetic field and temperature are $B=5\times 10^{12}\,{\rm G}$  and $T=4.31\,{\rm MeV}$.   The neutrino effective potential for $B\leq B_c=m^2/e\simeq$ 4.14 $\times10^{13}$ G  is given by \citep{2014arXiv1401.1581F}
{\small
\begin{eqnarray}
V_{eff}=\frac{\sqrt2 G_F \,m_e^3\,B}{\pi^2 B_c} \biggl[\sum^\infty_{l=0} (-1)^l \sinh\alpha_l\,\biggl\{\frac{m_e^2}{m^2_W}\biggl(1+4\frac{E^2_\nu}{m^2_e}\biggr)K_1(\sigma_l)\cr
+\sum^{\infty}_{n=1}\lambda_n\,\biggl( 2+\frac{m_e^2}{m^2_W}\biggl( 3-2\frac{B}{B_c}+4\frac{E^2_\nu}{m_e^2}\biggr)\biggr)\, K_1(\sigma_l\lambda_n) \biggr\}\cr \nonumber
-4\frac{m_e^2}{m^2_W}\frac{E_\nu}{m_e}\sum^\infty_{l=0} (-1)^l \cosh\alpha_l\biggl\{ \frac34K_0(\sigma_l)+\sum^{\infty}_{n=1}\lambda^2_n\, K_0(\sigma_l\lambda_n)  \biggr\}\biggr]\,
\label{fpoteff}
\end{eqnarray}
}
where K$_i$ is the modified Bessel function of integral order i and the parameters  as function of magnetic field (B), temperature (T)  and chemical potential ($\mu$) are  $\lambda^2_n=1+2\,n\,B/B_c$,    $\alpha_l=(l+1)\,\mu/T$ and $\sigma_l=(l+1)\,m_e/T$.
As shown in Fig. \ref{Veff}, the effective potential is an increasing function of magnetic field which has a value of $7.4 \times 10^{-10}$ eV for $B=5\times 10^{12}$ G. In this figure, we can observe that the effective potential is positive, therefore  due to the positivity of the effective potential ($V_{eff}>$ 0) neutrinos can oscillate resonantly. By considering two: solar  (left-hand figure above), atmospheric (right-hand figure above) and accelerator (left-hand figure below), and  three-neutrino (right-hand figure below) mixing, we have analyzed the resonance condition which is plotted in Fig. \ref{res_reg1}. In this figure, one can see that the temperature is a decreasing function of chemical potential independently of neutrino energy. In these plots we can see that temperature is a decreasing function of chemical potential and  for the values of temperature and chemical potential in the  range  $0.8\, {\rm MeV} \leq T \leq \,5\, {\rm MeV}$ and 10$^{-1}$ eV $\leq \mu \leq\,10^{3}$ eV, respectively, neutrinos oscillate resonantly. It can also be seen from these plots that the chemical potential achieves the largest value as accelerator parameters are considered, and the smallest one as solar parameters are taken into account.\\
In addition, as shown in fig \ref{pr_r1},  for the three-neutrino mixing we study the survival and conversion probabilities for the active-active ($\nu_{e,\mu,\tau} \leftrightarrow \nu_{e,\mu,\tau}$) neutrino oscillations as a function of  distance (left figure) and energy (right figure).  From these plots one can see that in the range considering distance and energy,  the probabilities of neutrino oscillation go from 0 to unity.
\subsection{Accretion shock region }
In this region, ($1.33\times 10^{6} {\rm cm} \leq r_2\leq r_s$) with $r_s$ given by eq. (\ref{rs}), the density profile can be obtained from the Chevalier model (eqs. \ref{equation1}, \ref{equation2} and \ref{equation3}) and be written as
\be\label{rho2}
\rho_2(r_2)=7.70\times 10^{2}\,\biggl(\frac{r_2}{r_s} \biggr)^{-3}~{\rm g~cm}^{-3}\,
\ee
\noindent where $r_2$ is the quasi-hydrostatic envelope radius.

\subsection{Free fall region}

The density  of material in free fall, $r_s\leq r_3 \leq r_h$  is obtained  from eqs. \ref{equation1} and written as
\be\label{rho3}
\rho_3(r)=5.74\times 10^{-2}\,\biggl(\frac{r_3}{r_h} \biggr)^{-3/2}~{\rm g~cm}^{-3}\,
\ee
where again $r_s$ is given by eq. (\ref{rs}) and  $r_h= 6.3\times 10^{10}\, {\rm cm}$.

\subsection{Surface of the star}
 
In particular, for the presupernova star of SN1987A (blue supergiant) a density profile was done by \citet{che89}. The analytic form of the density distribution in the outer radiative layer of the star has a polytropic structure, $\rho=\rho_0\left(R_{\star}/r-1\right)^{n}$, where $R_{\star}\simeq3\times10^{12}$ cm and $\rho_0\simeq3\times10^{-5}$ g cm$^{-3}$. The corresponding polytropic index for a radiative envelope with constant Thomson opacity is $n=3$.  \citet{che89} studied several models for the BSG presupernova of SN1987A, and all models are normalized to give the same density $\rho=2\left(10^{11}/10^{10.8}-1\right)^{3}=0.4$ g cm$^{-3}$. 
Then, the final density profile of the outer layer is given by a power-law fit with an index $n=17/7$,

\begin{eqnarray}
\rho_4(r)&=&3.4\times10^{-5}\mathrm{g\,cm^{-3}}\cr
&&\hspace{0.9cm}\times\cases{
\left(\frac{R_\star}{r}\right)^{17/7}; & $r_h< r < r_b$,\cr
\left(\frac{R_{\star}}{r}\right)^{17/7}\frac{\left(r-R_{\star}\right)^{5}}{\left(r_{b}-R_{\star}\right)^{5}}; & $r>r_b$.\cr
}
\label{rho4}
\end{eqnarray}
\noindent where $r_{b}=10^{12}$ cm.\\
Considering the density profiles  (eqs. \ref{rho2}, \ref{rho3} and \ref{rho4}), associating the number of electron per nucleon $Y_e=0.5$, then we obtain the number density of electrons as $N_e=N_A\,\rho(r) Y_e$ and the neutrino effective potential  $V_{eff}=\sqrt2 G_F N_e$. After that we present a description of two- and three-flavor neutrino oscillations.  From the resonance  condition,  we obtain the  resonance density ($\rho_R$)  as a function of resonance length ($l_R$) for two (eq. \ref{p1}) and three flavors (eq. \ref{p2}). We put together the plots of the density profiles as a function of distance and the resonance conditions (resonance density as a function of resonance length), as shown in fig \ref{res_reg234}.    For two flavors,   we have taken into account solar (left-hand figure above), atmospheric (right-hand figure above) and accelerator (left-hand figure below) parameters of neutrino experiments.   Using solar parameters, the resonance length is in the range  $\sim (6.1\times 10^6 -  2.3\times 10^{8})$ cm  and resonance density in $\sim  (1 - 10^{4})$ g/cm$^3$. For atmospheric parameters, the resonance length is less than $4.8\times 10^6$ cm and the resonance density in $\sim  (10^{2} - 10^{4.6})$ g/cm$^3$.   Using accelerator parameters, the resonance length is less than $4.3\times 10^5$ cm and  the resonance density lies in the range  $\sim  (10^{5} - 10^{6.8})$ g/cm$^3$ and  for three flavors (right-hand figure below),  the range of resonance length  is less than $\sim 4.2\times10^6$ cm  and resonance density is  $\sim  (10^{3.2} - 10^{4.7})$ g/cm$^3$.\\
In addition, as shown in figs. \ref{pr_r2}, \ref{pr_r3} and \ref{pr_r4}  for the three-neutrino mixing we study the survival and conversion probabilities for the active-active ($\nu_{e,\mu,\tau} \leftrightarrow \nu_{e,\mu,\tau}$) neutrino oscillations as a function of  distance (left-hand figure) and energy (right-hand figure).  From these plots one can see that  on one hand, the electron neutrino almost does not oscillate to any other flavor P$_{ee}\simeq$1, P$_{e\mu}\simeq$ 0 and P$_{e\tau}\simeq$ 0  and is almost independent of the energy of the neutrinos and the distance and on the other hand, the muon and tau neutrinos oscillate among themselves with equal probability and their oscillations depend on the neutrino energy and distance.\\
On the other hand, we calculate the flavor ratio expected on Earth for four neutrino energies  ($E_{\nu}=5$ MeV, 10 MeV, 15 MeV and 20 MeV), as shown in table \ref{flaratio_s}. From this table we can see a small deviation of the standard flavor 1:1:1 for neutrinos.   In this calculation we take into account that for neutrino cooling processes: electron-positron annihilation, inverse beta decay, nucleonic bremsstrahlung and plasmons, only  inverse beta decay is the one producing electron neutrino. It is important to say that our calculations of resonant oscillations were performed for neutrinos instead of anti-neutrinos, due to the positivity of the neutrino effective potential.
\begin{table}
\begin{center}\renewcommand{\tabcolsep}{0.2cm}
\renewcommand{\arraystretch}{1.1}
\begin{tabular}{ccccc}\hline
$E_{\nu}$  &$\phi_{\nu_e}:\phi_{\nu_\mu}:\phi_{\nu_\tau}$ &$\phi_{\nu_e}:\phi_{\nu_\mu}:\phi_{\nu_\tau}$&$\phi_{\nu_e}:\phi_{\nu_\mu}:\phi_{\nu_\tau}$  \\
(MeV)&On the Plasma &On the Star & On Earth\\ 
& (1.3$\times10^{6}$ cm)& (3$\times10^{12}$ cm)  &\\\hline
{\small 5}     &  {\small 0.914:1.042:1.042}  &  {\small 0.924:1.037:1.037} & {\small 0.977:1.007:1.015}  \\\hline

{\small 10}   & {\small 1.016:0.991:0.991}  &  {\small 1.013:0.993:0.993}&  {\small 1.004:0.998:0.997} \\\hline

{\small 15}  & {\small 1.106:0.947:0.946}  &  {\small 1.085:0.957:0.957} &    {\small 1.025:0.991:0.983} \\\hline

{\small 20}  &  {\small 1.145:0.927:0.927}  &   {\small 1.124:0.937:0.937}  &   {\small 1.037:0.987:0.975}  \\\hline

\end{tabular}
\end{center}
\caption{The flavor ratio on the surface of the plasma, star and Earth  for four neutrino energies  ($E_{\nu}=5$ MeV, 10 MeV, 15 MeV and 20 MeV).}
\label{flaratio_s}
\end{table}
\section{Discussion and Conclusions}  
\label{section5}
We have studied hypercritical accretion onto newborn NS with the aim of looking into the neutrino cooling effect in the formation of an envelope in quasi-hydrostatic equilibrium of SN1987A. 
This is particularly relevant in the context of making the NS eventually invisible as a pulsar following the supernova explosion, this with the condition that  the magnetic field is submerged by the accretion in the hypercritical regime.  Extending our exploration of parameter space to spherical symmetry spanning a significant fraction of the stellar surface in the simulations,  we have focused on the size of the neutrino-sphere where the emissivity of neutrinos is effective and then calculated the neutrino luminosity in the first instant after the reverse shock has reached the hard surface of the star and the formation of the quasi-steady atmosphere takes place.\\
We made comparisons of our numerical results with the analytical model of Chevalier and found an excellent agreement between the neutrino luminosity and height scale values obtained with the Flash Code and those estimated with analytical approximations. The additional neutrino processes involved in the numerical approach slightly increase the neutrino luminosity value and the height scale. This is because the 
analytical model of the hypercritical system was performed with some assumptions while the numerical approach has included more physical ingredients (magnetic field, a more realistic equation of state and additional neutrino processes).  We also analyze other interesting phenomena such as the submergence of the magnetic field on the stellar surface as well as the dynamics of the system with higher degrees of freedom.\\
On the other hand, taking into account the obtained neutrino luminosity and parameters of the SN-II, we estimated that 1.49 neutrinos must have reached the detector two hours later after the main neutrino burst.  However, because this number is much smaller that the $\sim 20$ events could have been undetected.   On the other hand, if considering the value of volume 31.9 kton in  SK-III  \citep{2012PhRvD..86a2006R}, we would expect a neutrino number of 22.3 in the hypercritical phase. The amount of these neutrinos would confirm such phase in other SN with the identical characteristics (temperature, distance, etc)
We have studied the active-active neutrino process in the supernova framework. We have divided the path of moving neutrinos into four regions.  For the first region, we have used the neutrino effective potential derived in \citet{2014arXiv1401.1581F} which is  a function of chemical potential ($\mu$), temperature ($T$),  neutrino energy ($E_\nu$)  and magnetic field.   We have shown that for a neutrino test of energy 1, 5 , 20 and 30 MeV, and parameters considered of temperature and chemical potential in the range of 1 MeV $\leq T \leq$ 5 MeV,   $10^{-2}\, {\rm eV} \leq \mu  \leq 10^3\, {\rm eV}$,  neutrinos oscillate resonantly, for two- and three-neutrino mixing.  In regions from two to four, we have also calculated each effective potential and then analyzed their oscillations through each zone.   For three neutrino mixing,  we have calculated the ratio flavor expected on Earth. Our analysis shows that deviations  from 1:1:1 are obtained  for neutrino energies of 5, 10, 15 and 20 MeV, given in table 1.  Distinct flavor ratios  will provide constraints on parameters of this hypercritical accretion.  Although currently it is very difficult to detect  neutrino oscillations, new techniques for detecting neutrino oscillations arise in the near future allowing us to confirm our estimates and thus the hypercritical phase.\\
It is important to highlight that we estimate a detectable number of neutrinos assuming an object at the distance of SN 1987A at the LMC, however it could be used for future predictions in a more favorable situation where a core collapse might take place in the Milky Way Galaxy. 

\section*{Acknowledgements}
We are grateful to DGTIC-UNAM and to IA-UNAM for allowing us to use their NES and ATOCATL Clusters where all the simulations were performed. The software used in this work was in part developed by the DOE NNSA-ASC OASCR Flash Center at the University of Chicago. This work was supported in part by CONACyT grants CB-2009-1 No. 132400, CB-2008-1 No. 101958, project 128556-F and project 165584. Also was supported by PAPIIT project IN106212. Also we thank to Dany Page and William Lee for useful discussions. This work was supported by Luc Binette scholarship and the projects IG100414 and Conacyt 101958.


%
\clearpage
\begin{figure*}[htp]\label{res_reg1}
\centering
\includegraphics[width=\textwidth]{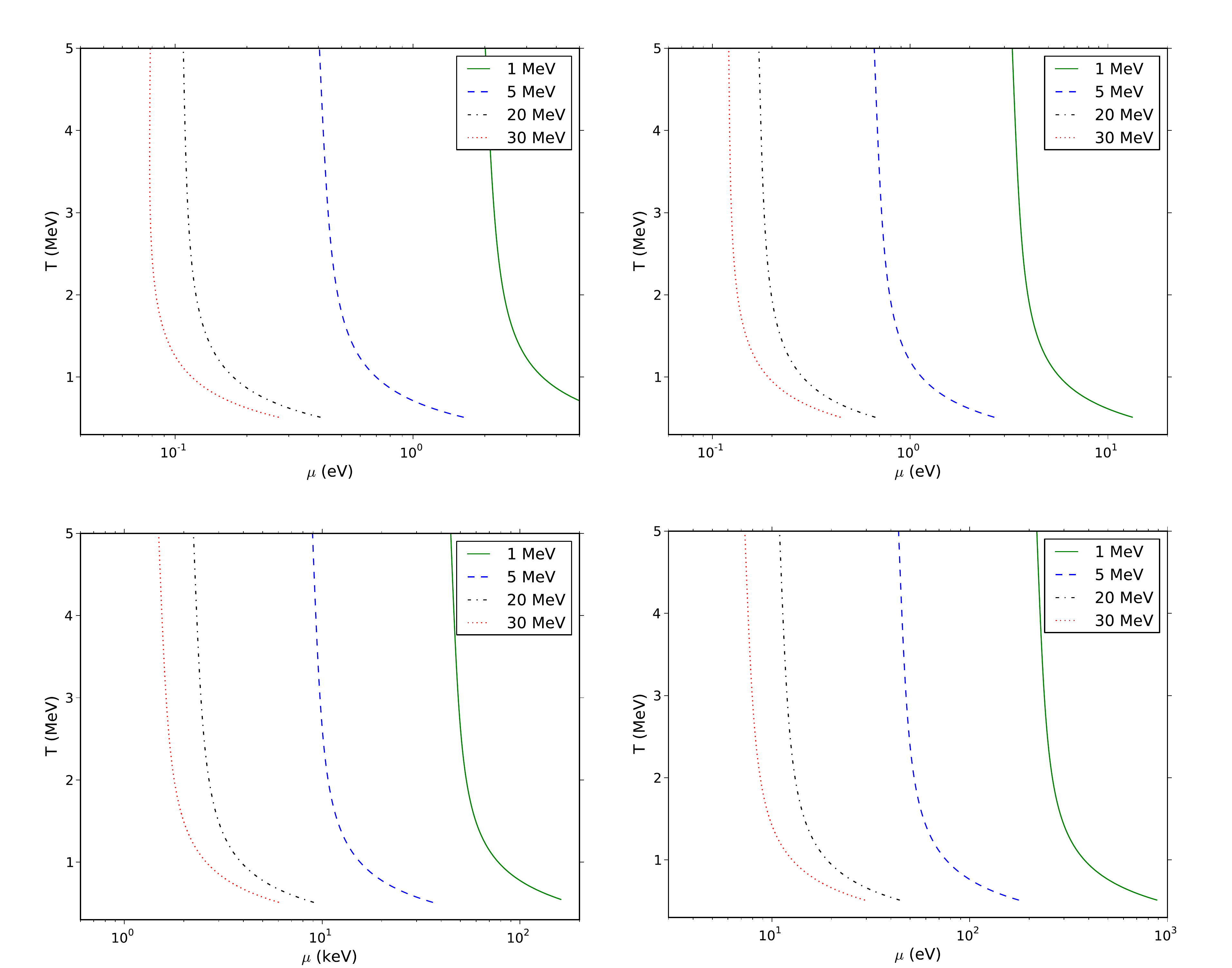}
\caption{\label{res_reg1}  Plot of temperature  ($T$) as a function of chemical potential  ($\mu$) for which the resonance condition is satisfied.  We have used the best fit parameters of the two-flavor: solar (left-hand figure above), atmospheric (right-hand figure above) and accelerator (left-hand figure below), and three-flavor (right-hand figure below)  neutrino oscillation, the value of magnetic field of  $B=5\times 10^{12}$ G and taken into account four  different neutrino energies:  E$_\nu=$1 MeV  (green thin-solid line) (red dashed line),  E$_\nu=$5 MeV (blue dashed line), E$_\nu=20$ MeV (black  dot-dashed line) and  E$_\nu=30$ MeV (bred dotted line)). }
\end{figure*}
\begin{figure*}[htp]\label{flavors}
\centering
\includegraphics[width=\textwidth]{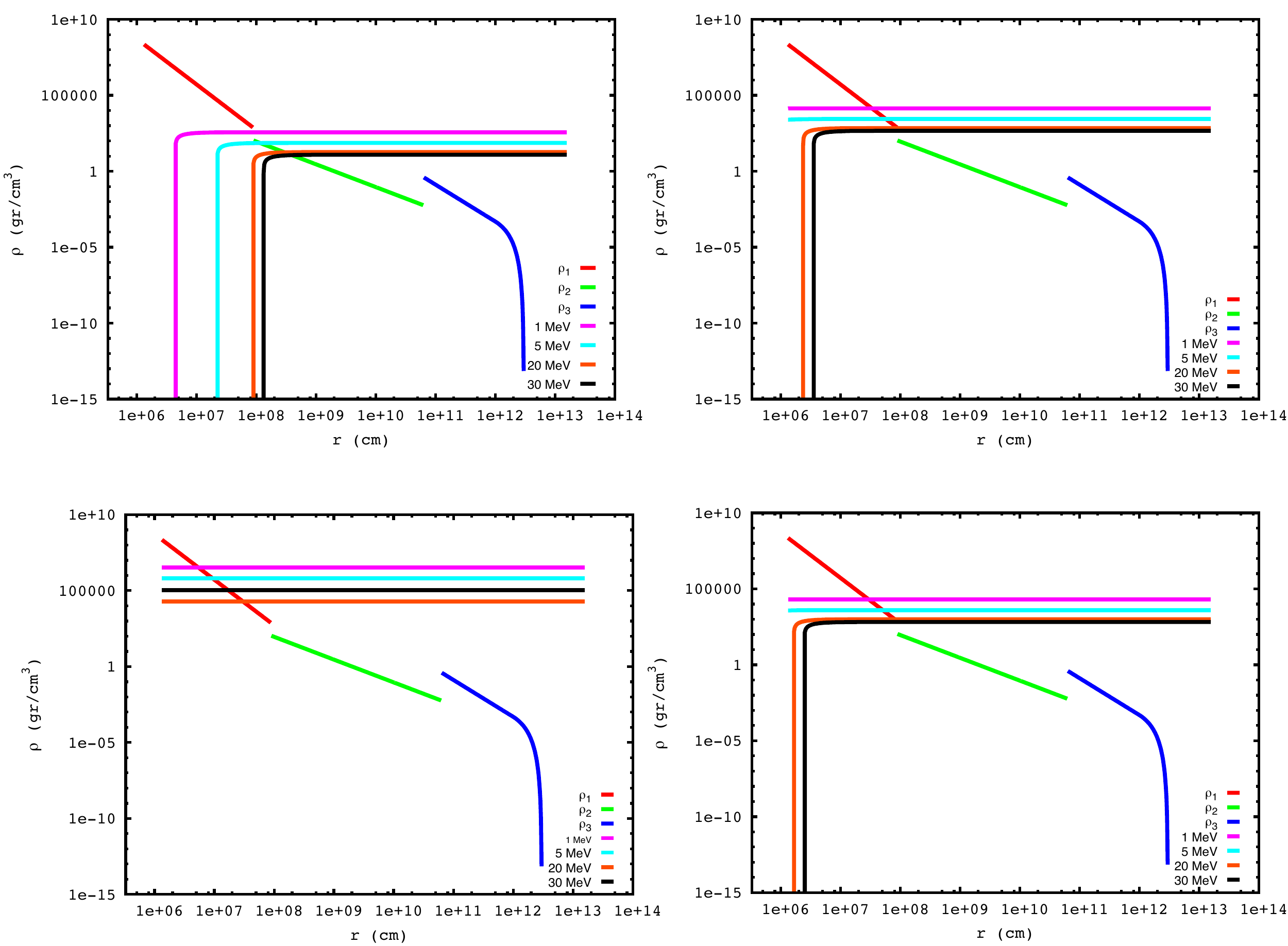}
\caption{\label{res_reg234}  Density profiles  ($\rho_2$, $\rho_3$ and $\rho_4$)  given by eqs. (\ref{rho2}), (\ref{rho3}) and (\ref{rho4}), respectively are plotted.  Also from the resonance condition,  we plot  the resonance density as a function of resonance length for High-energy neutrinos. We have used the best fit parameters  of the two-flavor: solar (left-hand figure above), atmospheric (right-hand figure above) and accelerator (left-hand figure below), and three-flavor (right-hand figure below)  neutrino oscillation.}
\end{figure*}
\begin{figure*}[htp]
\centering
\includegraphics[width=\textwidth]{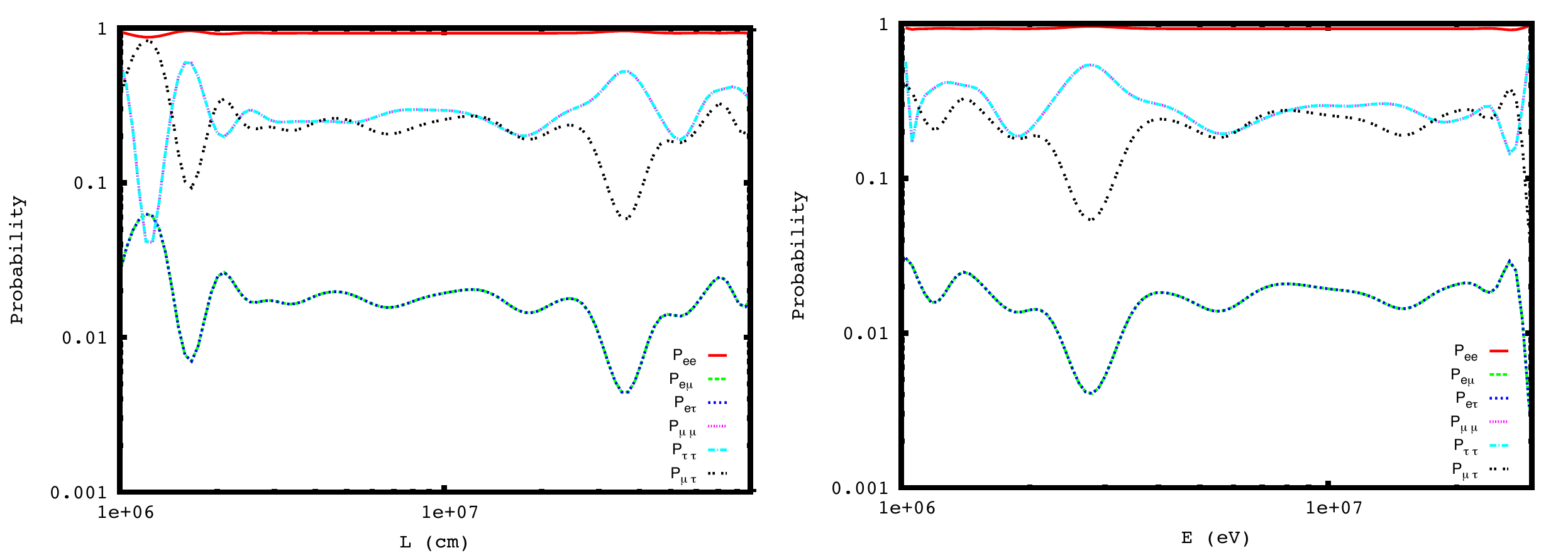}
\caption{\label{pr_r2} We plot the oscillation probability as a function of distance (left figure) and energy (right figure) when neutrinos are propagating through region 2.}
\end{figure*}
\begin{figure*}[htp]
\centering
\includegraphics[width=\textwidth]{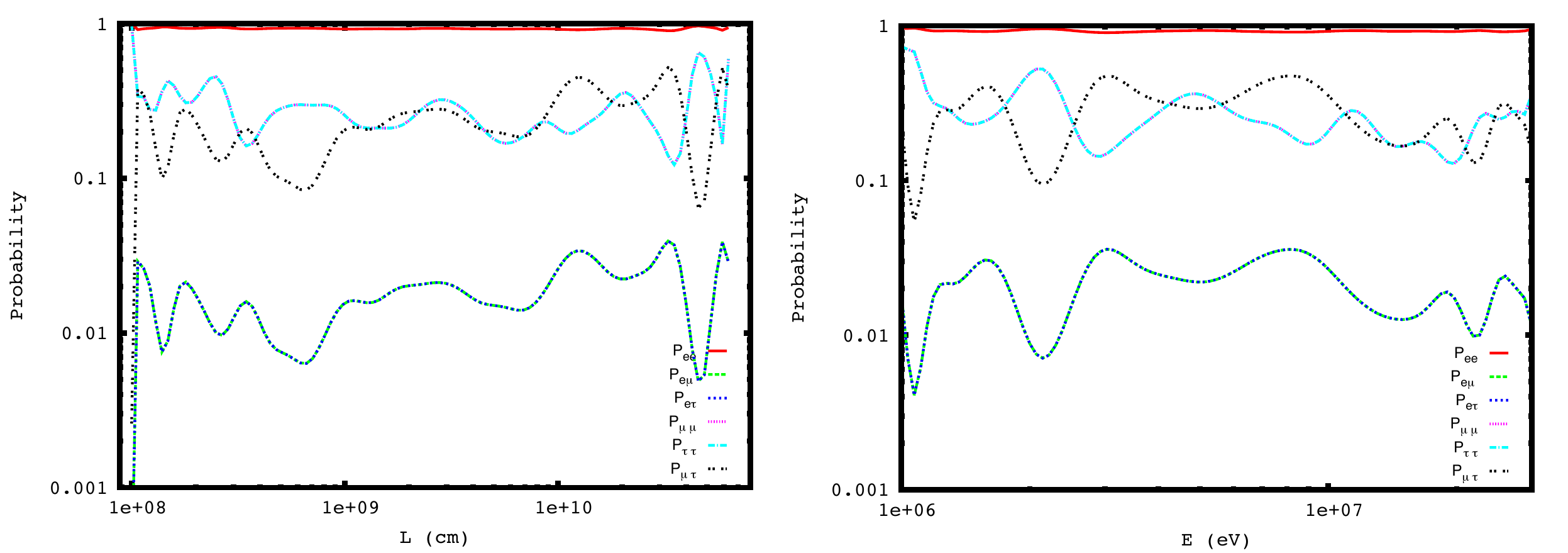}
\caption{\label{pr_r3} We plot the oscillation probability as a function of distance (left figure) and energy (right figure) when neutrinos are propagating through region 3.}
\end{figure*}
\begin{figure*}[htp]
\centering
\includegraphics[width=\textwidth]{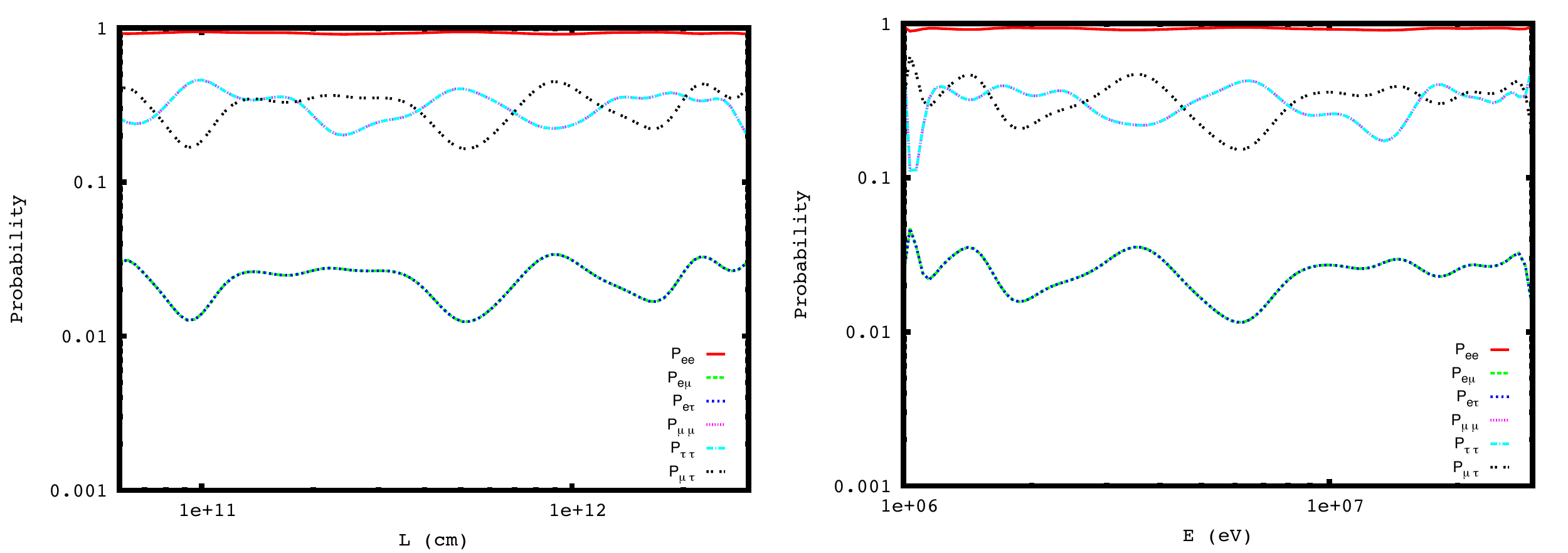}
\caption{\label{pr_r4} We plot the oscillation probability as a function of distance (left figure) and energy (right figure) when  neutrinos are propagating through region 4.}
\end{figure*}
\end{document}